\documentclass[pra,aps,superscriptaddress]{revtex4}

\usepackage{amsmath,amsfonts,amssymb,caption,hyperref,color,epsfig,graphics,graphicx,latexsym,mathrsfs,revsymb,theorem,url,verbatim,epstopdf}
\usepackage{subfigure}
\hypersetup{colorlinks,linkcolor={blue},citecolor={blue},urlcolor={red}}

\usepackage{hyperref}

\makeatletter

\newcommand{\Rmnum}[1]{\expandafter\@slowromancap\romannumeral #1@}
\makeatother
\newtheorem{definition}{Definition}
\newtheorem{proposition}[definition]{Proposition}

\newtheorem{conjecture}[definition]{Conjecture}

\newtheorem{remark}[definition]{Remark}
\newtheorem{example}[definition]{Example}
\newtheorem{question}[definition]{Question}

\def\squareforqed{\hbox{\rlap{$\sqcap$}$\sqcup$}}
\def\qed{\ifmmode\squareforqed\else{\unskip\nobreak\hfil
		\penalty50\hskip1em\null\nobreak\hfil\squareforqed
		\parfillskip=0pt\finalhyphendemerits=0\endgraf}\fi}
\def\endenv{\ifmmode\;\else{\unskip\nobreak\hfil
		\penalty50\hskip1em\null\nobreak\hfil\;
		\parfillskip=0pt\finalhyphendemerits=0\endgraf}\fi}
\newenvironment{proof}{\noindent \textbf{{Proof.~} }}{\qed}
\def\Dbar{\leavevmode\lower.6ex\hbox to 0pt
	{\hskip-.23ex\accent"16\hss}D}
\makeatletter
\def\url@leostyle{%
	\@ifundefined{selectfont}{\def\UrlFont{\sf}}{\def\UrlFont{\small\ttfamily}}}
\makeatother
\allowdisplaybreaks

\def\bcj{\begin{conjecture}}
	\def\ecj{\end{conjecture}}
\def\bcr{\begin{corollary}}
	\def\ecr{\end{corollary}}
\def\bd{\begin{definition}}
	\def\ed{\end{definition}}
\def\bea{\begin{eqnarray}}
\def\eea{\end{eqnarray}}
\def\bem{\begin{enumerate}}
	\def\eem{\end{enumerate}}
\def\bex{\begin{example}}
	\def\eex{\end{example}}
\def\bim{\begin{itemize}}
	\def\eim{\end{itemize}}
\def\bl{\begin{lemma}}
	\def\el{\end{lemma}}
\def\bma{\begin{bmatrix}}
	\def\ema{\end{bmatrix}}
\def\bpf{\begin{proof}}
	\def\epf{\end{proof}}
\def\bpp{\begin{proposition}}
	\def\epp{\end{proposition}}
\def\bqu{\begin{question}}
	\def\equ{\end{question}}
\def\br{\begin{remark}}
	\def\er{\end{remark}}
\def\bt{\begin{theorem}}
	\def\et{\end{theorem}}

\def\btb{\begin{tabular}}
	\def\etb{\end{tabular}}

\newcommand{\nc}{\newcommand}

\def\i{\iota}

\def\r{\rho}

\nc{\bbA}{\mathbb{A}} \nc{\bbB}{\mathbb{B}} \nc{\bbC}{\mathbb{C}}
\nc{\bbD}{\mathbb{D}} \nc{\bbE}{\mathbb{E}} \nc{\bbF}{\mathbb{F}}
\nc{\bbG}{\mathbb{G}} \nc{\bbH}{\mathbb{H}} \nc{\bbI}{\mathbb{I}}
\nc{\bbJ}{\mathbb{J}} \nc{\bbK}{\mathbb{K}} \nc{\bbL}{\mathbb{L}}
\nc{\bbM}{\mathbb{M}} \nc{\bbN}{\mathbb{N}} \nc{\bbO}{\mathbb{O}}
\nc{\bbP}{\mathbb{P}} \nc{\bbQ}{\mathbb{Q}} \nc{\bbR}{\mathbb{R}}
\nc{\bbS}{\mathbb{S}} \nc{\bbT}{\mathbb{T}} \nc{\bbU}{\mathbb{U}}
\nc{\bbV}{\mathbb{V}} \nc{\bbW}{\mathbb{W}} \nc{\bbX}{\mathbb{X}}
\nc{\bbZ}{\mathbb{Z}}

\nc{\bA}{{\bf A}} \nc{\bB}{{\bf B}} \nc{\bC}{{\bf C}}
\nc{\bD}{{\bf D}} \nc{\bE}{{\bf E}} \nc{\bF}{{\bf F}}
\nc{\bG}{{\bf G}} \nc{\bH}{{\bf H}} \nc{\bI}{{\bf I}}
\nc{\bJ}{{\bf J}} \nc{\bK}{{\bf K}} \nc{\bL}{{\bf L}}
\nc{\bM}{{\bf M}} \nc{\bN}{{\bf N}} \nc{\bO}{{\bf O}}
\nc{\bP}{{\bf P}} \nc{\bQ}{{\bf Q}} \nc{\bR}{{\bf R}}
\nc{\bS}{{\bf S}} \nc{\bT}{{\bf T}} \nc{\bU}{{\bf U}}
\nc{\bV}{{\bf V}} \nc{\bW}{{\bf W}} \nc{\bX}{{\bf X}}
\nc{\bZ}{{\bf Z}}

\nc{\cA}{{\cal A}} \nc{\cB}{{\cal B}} \nc{\cC}{{\cal C}}
\nc{\cD}{{\cal D}} \nc{\cE}{{\cal E}} \nc{\cF}{{\cal F}}
\nc{\cG}{{\cal G}} \nc{\cH}{{\cal H}} \nc{\cI}{{\cal I}}
\nc{\cJ}{{\cal J}} \nc{\cK}{{\cal K}} \nc{\cL}{{\cal L}}
\nc{\cM}{{\cal M}} \nc{\cN}{{\cal N}} \nc{\cO}{{\cal O}}
\nc{\cP}{{\cal P}} \nc{\cQ}{{\cal Q}} \nc{\cR}{{\cal R}}
\nc{\cS}{{\cal S}} \nc{\cT}{{\cal T}} \nc{\cU}{{\cal U}}
\nc{\cV}{{\cal V}} \nc{\cW}{{\cal W}} \nc{\cX}{{\cal X}}
\nc{\cZ}{{\cal Z}}

\nc{\hA}{{\hat{A}}} \nc{\hB}{{\hat{B}}} \nc{\hC}{{\hat{C}}}
\nc{\hD}{{\hat{D}}} \nc{\hE}{{\hat{E}}} \nc{\hF}{{\hat{F}}}
\nc{\hG}{{\hat{G}}} \nc{\hH}{{\hat{H}}} \nc{\hI}{{\hat{I}}}
\nc{\hJ}{{\hat{J}}} \nc{\hK}{{\hat{K}}} \nc{\hL}{{\hat{L}}}
\nc{\hM}{{\hat{M}}} \nc{\hN}{{\hat{N}}} \nc{\hO}{{\hat{O}}}
\nc{\hP}{{\hat{P}}} \nc{\hR}{{\hat{R}}} \nc{\hS}{{\hat{S}}}
\nc{\hT}{{\hat{T}}} \nc{\hU}{{\hat{U}}} \nc{\hV}{{\hat{V}}}
\nc{\hW}{{\hat{W}}} \nc{\hX}{{\hat{X}}} \nc{\hZ}{{\hat{Z}}}

\nc{\hn}{{\hat{n}}}

\def\max{\mathop{\rm max}}
\def\min{\mathop{\rm min}}

\def\tr{\mathop{\rm Tr}}

\newcommand{\bra}[1]{\langle#1|}
\newcommand{\ket}[1]{|#1\rangle}

\begin{document}
\title{Evolution of Quantum Resources in Quantum-walk-based Search Algorithm}
\author{Meng Li}
\email[]{limeng2021@ict.ac.cn}
\affiliation{State Key Lab of Processors, Institute of Computing Technology, Chinese Academy of Sciences, Beijing 100190, China}
\author{Xian Shi}
\email[]{shixian01@gmail.com (corresponding author)}
\affiliation{College of Information Science and Technology, Beijing University of Chemical Technology, Beijing 100029, China}

\begin{abstract}{Quantum walk is fundamental to designing many quantum algorithms. Here we consider the effects of quantum coherence and quantum entanglement for the quantum walk search on the complete bipartite graph. 
First, we numerically show the complementary relationship between the success probability and the two quantum resources (quantum coherence and quantum entanglement). We also provide theoretical analysis in the asymptotic scenarios. At last, we discuss the role played by generalized depolarizing noises and find that it would influence the dynamics of success probability and quantum coherence sharply, which is demonstrated by theoretical derivation and numerical simulation.}
\end{abstract}

\maketitle

\section{Introduction}
\indent Compared with the classical random walk, quantum walk spreads quadratically faster on the line \cite{ambainis2001one} and mixes quadratically faster on the cycle \cite{aharonov2001quantum}.
Thus, quantum walks have been used widely in algorithm design \cite{aaronson2003quantum, ambainis2007quantum}.
As a universal quantum computing model\cite{childs2009universal, childs2013universal, lovett2010universal, underwood2010universal}, physical implementations of quantum walk have been realized in various experimental platforms, such as ion traps \cite{schmitz2009quantum}, optical \cite{tang2018experimental}, and superconducting processors \cite{gong2021quantum}.

Thanks to the particularity and parallelism of quantum nature itself, quantum walk has large application ranges and application potentials. Of all the quantum resources, quantum coherence and entanglement are different from the classical world significantly \cite{chitambar2019quantum}. One of the most important problems in quantum resource theories is how to quantify them \cite{chitambar2019quantum,horodecki2009quantum,streltsov2017colloquium}. The problem of entanglement owns a history of more than twenty years, 
and it still attracts much attention up to now. Bennett $et$ $al.$ presented entanglement measures with operational significance \cite{bennett1996mixed}.  In 1997, Vedral $et$ $al.$ presented necessary conditions for an entanglement measure \cite{vedral1997quantifying}, Vidal considered the stronger conditions for an entanglement measure and presented a method to build  \cite{vidal2000entanglement}. Other than the generic building method of entanglement measures, some meaningful thoughts to build the entanglement measures are presented. The robustness of an entangled state quantifies the minimal
mixing required to destroy all the entanglement of the state \cite{vidal1999robustness}. In \cite{gour2020optimal,shi2021extension}, the authors presented and investigated  the entanglement measures building with operational  meanings. Moreover, the entanglement measures can also be defined in terms of the geometrical ways, such as the geometric measure of entanglement \cite{wei2003geometric}, the entanglement measures based on the quantum relative entropy \cite{vedral1997quantifying} and 1-norm distance \cite{shi2022quantifying}. The ideas of many measures on coherence are based on the entanglement measures, such as, 1-norm coherence \cite{rana2016trace}, geometric measure of coherence \cite{streltsov2015measuring}. Comparing with the above two measures, $l_1$ norm of coherence is easier to evaluate \cite{baumgratz2014quantifying}. Here we use the $l_1$ norm of coherence to quantify the resource's evolution.

Quantum coherence and quantum entanglement play important roles in the specific dynamical processes of quantum walk.
Maloyer $et$ $al.$ analyzed the entanglement between coin space and position space of the walker state numerically and further considered the optimal decoherence rate \cite{maloyer2007decoherence}.
Rodriguez $et$ $al.$ discussed the quantum discord and entanglement between two coin states of the two-particle quantum walk on cycle graphs  \cite{rodriguez2015discord}.
For the quantum walk on cycles, He $et$ $al.$ also analyzed the dynamics of the $l_{1}$ norm coherence and the corresponding influences of unitary noises \cite{he2017coherence}. All the above analyses of quantum resources describe the properties of quantum walk itself on different graphs.

Besides, quantum coherence and entanglement are directly related to the performance of various algorithms and even provide potential power in many kinds of quantum algorithms.
Grover search algorithm (GSA), as one of the landmark results of the golden age of quantum computing, demonstrated the quadratic speed-up advantage over classical algorithms.
In the view of concurrence, Fang $et$ $al.$ focused on the degree of entanglement and pointed out that it has some relationship with the success probability to some extent \cite{fang2005entanglement},
and Rungta explained the optimality of the GSA by giving a necessary and sufficient condition for quadratic speed-up from the entanglement perspective \cite{rungta2009quadratic}.
Using the geometric measure of entanglement, the dynamics of entanglement under the iterations in the GSA have also been described \cite{chakraborty2013entanglement, pan2017global}.
Besides, each operator in GSA, such as the oracle operator and reflection operator, plays different roles in the entanglement and coherence dynamics \cite{pan2019operator, pan2019entangling}. 
The success probability of Grover algorithm also depends on quantum coherence depletion in terms of both the relative entropy and $l_{1}$ norm of coherence measure \cite{shi2017coherence}.
Also, there is a complementary relationship between success probability and quantum coherence in GSA \cite{pan2022complementarity}.

Noise can affect the performance of quantum algorithms, such as success probability, efficiency and time. In 2000, Long $et$ $al.$ considered the influences of
different imperfect gate operations of the quantum search algorithm in the absence of decoherence and error corrections \cite{long2000dominant}. Shenvi $et$ $al.$ examined the robustness of GSA to a random phase error in the oracle and analyzed the complexity of the search process \cite{shenvi2003effects}. Gawron $et$ $al.$ studied the influence of noise on the computational complexity of the GSA \cite{gawron2011noise}. In 2019, Reitzner and Hiller studied GSA under the influence of localized partially dephasing noise of rate 
$p$ \cite{reitzner2019grover}. Recently, Wang $et$ $al.$ addressed a series of simulations by inflicting various types of noise, modeled by the IBM qiskit \cite{wang2020prospect}. The above works studied the GSA under noise theoretically and experimentally, nevertheless, they did not present the changes with the quantum correlations. Recently, Rastegin and his coauthor studied the case when the oracle of the GSA is exposed to noise. They also analyzed the relationship between coherence and success probability \cite{rastegin2018degradation,rastegin2022quantum}.

In fact, the well-known Grover algorithm is essentially a quantum walk search on the complete graph. Also, the relationship between its performance and quantum resources has been analyzed and characterized. A natural question is what happens on a complete bipartite graph.
{It is of interest to know whether these properties and conclusions still hold true in the case of complete bipartite graph. In fact, the problem of searching on complete bipartite graph is fundamental and important. Since this search algorithm was proposed \cite{rhodes2019quantum}, it has been widely concerned by researchers \cite{rapoza2021search,xu2022robust}.}
Hence here we aim to analyze the relationship between quantum resources, including quantum coherence and entanglement, and the performance of quantum-walk-based search on the complete bipartite graph. 
Furthermore, we explore the evolution of success probability and quantum coherence under generalized depolarizing noises.

This paper is organized as follows. In section \ref{sec1}, we briefly introduce the model of quantum walk search. In section \ref{sec2}, we investigate the evolution of quantum resources in the quantum-walk-based algorithm and its relationship with the success probability. Specifically, the evolution of quantum coherence and entanglement will be shown in section \ref{ssec1} and \ref{ssec2}, respectively. In section \ref{ssec3}, we discuss the dynamics of quantum coherence under generalized depolarizing noises. At last, we end with a discussion and outlook in section \ref{sum}. 

\section{Preliminary}\label{sec1}
Given a graph $G(V,E)$, where $V$ is a vertex set of size $N$, and $E$ is an edge set.
The Hilbert space in which quantum walks evolve is spanned by the computational basis $|i,j\rangle$, where $|i,j\rangle$ means a particle at vertex $i$ pointing to $j$.
One step of quantum walk is $$W=S(I\otimes C),$$ where $S=\sum_{(i,j)\in E}|i,j\rangle\langle j,i|$ is the conditional shift operator and $C$ is the coin operator acting on coin spaces. 
When it comes to the search problem, there is an oracle $Q$ that flips the sign of the amplitude of the marked vertices. Thus the quantum walk based search operator can be described as $$U=WQ=S(I\otimes C)Q.$$ 
\par
\begin{figure} [htbp]
    \centering
    \includegraphics[width=50mm]{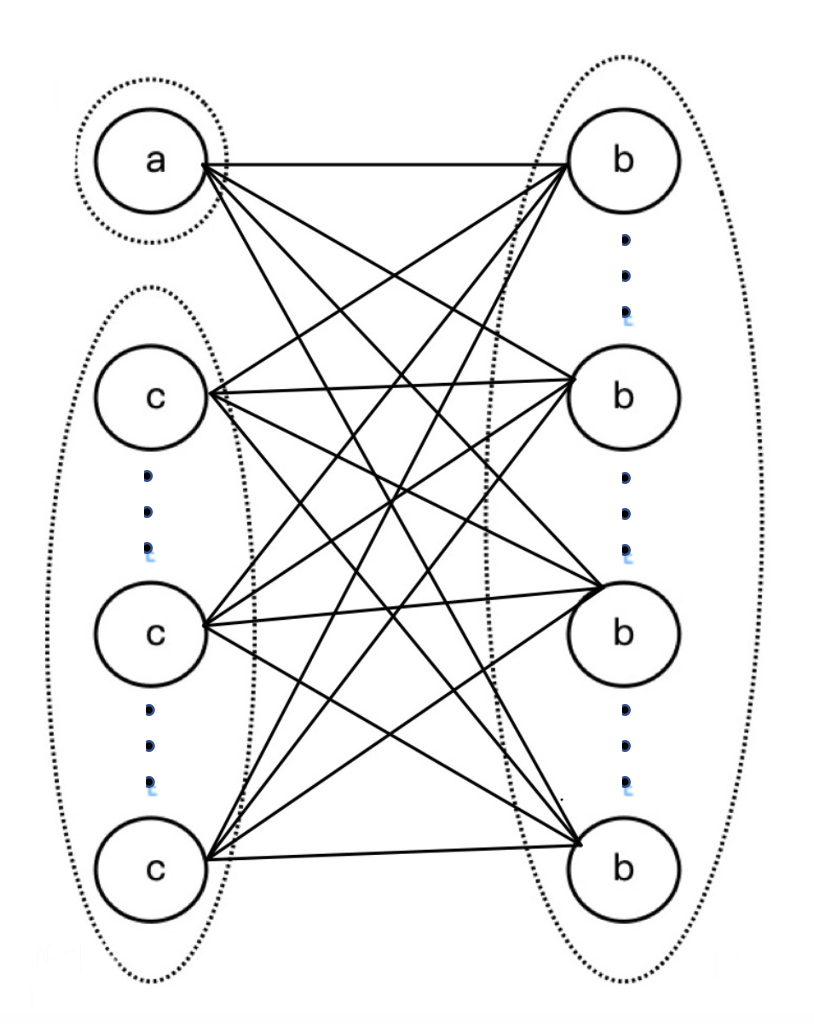}
    \caption{A special case on the complete bipartite graph when $N_1=N_2$ and k=1.} 
    \label{1}
\end{figure}

 In \cite{rhodes2019quantum},  the authors investigate the algorithm on searching the marked vertices in the complete bipartite graph of N vertices with two initial states. The complete bipartite graph owns $N_{1}$ and $N_{2}$ vertices in partite $X$ and $Y$, respectively, where $N_{1}+N_{2}=N$ and $X\cup Y=V$. {In the following, we denote $a,c$ and $b$ as the marked vertices, unmarked vertices in set $X$ and vertices in set $Y$, respectively, and there are $k$ marked vertices in set $X$.} An example of the complete bipartite graph is plotted in Fig. \ref{1}. According to the vertex-edge connection relation, the complete bipartite graph can be reduced to a invariant subspace for discussion which will simplify the calculation and analysis. The subspace can be spanned by $\{|ab\rangle, |ba\rangle, |bc\rangle, |cb\rangle\}$, where { $\ket{a}=\frac{1}{\sqrt{k}}\sum_{a}\ket{a},\ket{b}=\frac{1}{\sqrt{N_2}}\sum_{b}\ket{b}$, and $\ket{c}=\frac{1}{\sqrt{N_1-k}}\sum_{c}\ket{c}$.} 
{Furthermore, the initial state should contain no information about the marked vertices, and the probability of each vertex should be equal.}
The initial states here we consider are written in the following forms:
\begin{align}
|s\rangle=\frac{1}{\sqrt{N_{1}+N_{2}}}(\sqrt{k}|ab\rangle+\sqrt{\frac{N_{2}k}{N_{1}}}|ba\rangle+\sqrt{\frac{N_{2}(N_{1}-k)}{N_{1}}}|bc\rangle+\sqrt{N_{1}-k}|cb\rangle),\label{is1}\\
 |\sigma\rangle=\frac{1}{\sqrt{2N_{1}N_{2}}}(\sqrt{kN_{2}}(|ab\rangle+|ba\rangle)+\sqrt{N_{2}(N_{1}-k)}(|bc\rangle+|cb\rangle)) \label{is2}
\end{align}
{The initial state $|s\rangle$ is a common choose, which is an equal superposition state of all the vertices. So if the position space is measured, the probability of measuring each vertex is equal. Also, when the given graph is regular, we have $W|s\rangle=|s\rangle$ which means the walk operator itself does not yield any information about the marked vertices. However, for the complete bipartite graph with $N_{1}\neq N_{2}$, this property does not hold. Thus, here we would also consider the initial state $|\sigma\rangle$ that is an equal superposition state of all the edges. For the initial state $|\sigma\rangle$, it satisfies the property $W|\sigma\rangle=|\sigma\rangle$ and the probability of measuring each edge is equal.}

Next we give the final state in computational basis  for the quantum walk search on complete bipartite graph.
When the initial state is $|s\rangle$,
the final state can be written as
\begin{equation}\label{ts1}
\ket{\psi(t)}=\left\{\begin{aligned}
&\cos\theta[\sin(\delta+\phi t)\ket{ab}+\cos(\delta+\phi t)\ket{cb}]+\sin\theta[\sin(\delta-\phi t)\ket{ba}+\cos(\delta-\phi t)\ket{bc}]\quad t=2k,k\in\mathbb{N}^{+}\\
&\sin\theta[\sin(\phi(1+t)-\delta)\ket{ab}+\cos(\phi(1+t)-\delta)\ket{cb}]\\+&\cos\theta[\sin((1-t)\phi-\delta)\ket{ba}+\cos((t-1)\phi+\delta)\ket{bc}]\quad t=2k-1,k\in\mathbb{N}^{+}
\end{aligned}\right.
\end{equation}
{The corresponding success probability is
\begin{equation}\label{tp1}
P_{t}=\left\{\begin{aligned}
&\cos^{2}\theta\sin^{2}(\delta+\phi t)\quad t=2k,k\in\mathbb{N}^{+}\\
&\sin^{2}\theta\sin^{2}(\phi(1+t)-\delta)\quad t=2k-1,k\in\mathbb{N}^{+}
\end{aligned}\right.
\end{equation}}

When the initial state is $|\sigma\rangle$, the final state can be written as
\begin{equation}\label{ts2}
\ket{\tilde{\psi}(t)}=\left\{\begin{aligned}
&\frac{\sqrt{2}}{2}(\sin(\phi t+\delta)\ket{ab}+\sin(\delta-\phi t)\ket{ba}+\cos(\phi t-\delta)\ket{bc}+\cos(\phi t+\delta)\ket{cb})\quad t=2k,k\in\mathbb{N}^{+}\\
&\frac{\sqrt{2}}{2}(\sin(\phi t+\phi-\delta)\ket{ab}+\sin(\phi-\delta-\phi t)\ket{ba}\\+&\cos(\phi t-\phi+\delta)\ket{bc}+\cos(\phi t+\phi-\delta)\ket{cb})\quad t=2k-1,k\in\mathbb{N}^{+}
\end{aligned}\right.
\end{equation}
{The success probability in this scenario can be described as \begin{equation}
\tilde{P}_{t}=\left\{\begin{aligned}
&\frac{1}{2}\sin^{2}(\phi t+\delta)\quad t=2k,k\in\mathbb{N}^{+}\\
&\frac{1}{2}\sin^{2}(\phi t+\phi-\delta)\quad t=2k-1,k\in\mathbb{N}^{+}
\end{aligned}\right.\label{sp}
\end{equation}}
Here $\theta,\delta$ and $\phi$ in $(\ref{ts1})-(\ref{sp})$ are defined as $\theta=\arccos\frac{\sqrt{N_1}}{\sqrt{N_1+N_2}},\delta=\arcsin\sqrt{\frac{k}{N_1}},\phi=\arccos{\sqrt{\frac{N_{1}-k}{N_{1}}}},$ respectively.\par

\section{The Effects of Quantum Resources on this Algorithm} \label{sec2}
\indent In this section, we mainly consider the evolution of two kinds of quantum resources, coherence and entanglement during the iterations of the quantum-walk-based search algorithms. In fact, the performance of the algorithm is directly related to the quantum resource.
Furthermore, we show there exists strong relationships between quantum resources and success probability, such as maximum (minimum) relativity, periodicity and complementarity.\par
\subsection{Dynamics of Coherence in the Algorithm Progress}\label{ssec1}
Quantum coherence is one of the important components of quantum information theory, and its role in quantum communication protocols and quantum algorithms has attracted more and more attention.
Given a density matrix $\rho$, the $l_{1}$ norm coherence of $\rho$ can be described as:
\begin{align}
    C_{l_{1}}(\rho)=\sum_{i\neq j}|\rho_{ij}|=\sum_{i\neq j}|\langle i|\rho|j\rangle|,\label{coherence}
\end{align}
where $\{|i\rangle\}$ is the given basis of the whole Hilbert space. And the upper bound of the $l_{1}$ norm coherence for any quantum state in a $d$-dimensional Hilbert system is $d-1$. 
Thus, we call $\frac{C_{l_{1}}(\rho)}{d-1}$ the normalized coherence of $\rho$, {which has also been considered in Ref. \cite{pan2022complementarity}.} For the initial state (\ref{is1}) and the state after $t$-th step quantum walk (\ref{ts1}), we can calculate the quantum coherence {in the computational basis $\{|ab\rangle, |ba\rangle, |bc\rangle, |cb\rangle$\}} by its definition (\ref{coherence}).
We have
\begin{equation}
\begin{split}
    C_{l_{1}}(\ket{\Psi(t)})=&
    \sin{2\theta}\left(|\sin{(A+h\phi)}|+|\cos{(A+h\phi)}|\right)\left(|\sin{(B+h\phi)}|+|\cos{(B+h\phi)}|\right)\\
    &+\cos^{2}{\theta}|\sin{2(A+h\phi)}|+\sin^{2}{\theta}|\sin{2(B+h\phi)}|,
\end{split}
\end{equation}
where $A=\delta+t\phi$, $B=\delta-t\phi$, $h=\frac{(-1)^t-1}{2}$.
{The corresponding success probability is \begin{equation}
P_{t}=\left\{\begin{aligned}
&\cos^{2}\theta\sin^{2}A\quad t=2k,k\in\mathbb{N}^{+}\\
&\sin^{2}\theta\sin^{2}(B-\phi)\quad t=2k-1,k\in\mathbb{N}^{+}
\end{aligned}\right.
\end{equation}}
And for the $Case$ (ii), we have 
\begin{equation}
\begin{split}
    C_{l_{1}}(|\tilde{\Psi}(t)\rangle)=&\left(|\sin{(A+h\phi)}|+|\cos{(A+h\phi)}|\right)\left(|\sin{(B+h\phi)}|+|\cos{(B+h\phi)}|\right)\\
    &+\frac{|\sin{(2(A+h\phi))}|+|\sin{(2(B+h\phi))}|}{2},
\end{split}
\end{equation}
which can be obtained similarly.
{The corresponding success probability can be represented as \begin{equation}
\tilde{P}_{t}=\left\{\begin{aligned}
&\frac{1}{2}|\sin{A}|^{2}\quad t=2k,k\in\mathbb{N}^{+}\\
&\frac{1}{2}|\sin{(B-\phi))}|^{2}\quad t=2k-1,k\in\mathbb{N}^{+}
\end{aligned}\right.
\end{equation}}

\begin{figure}[htbp]
    \centering    
    \subfigure{\includegraphics[width=85mm]{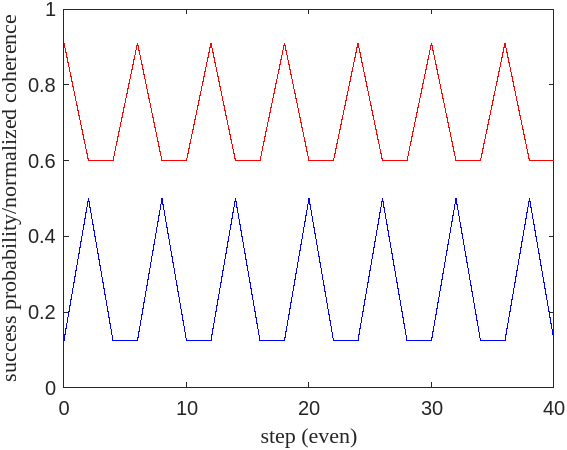}} \subfigure{\includegraphics[width=85mm]{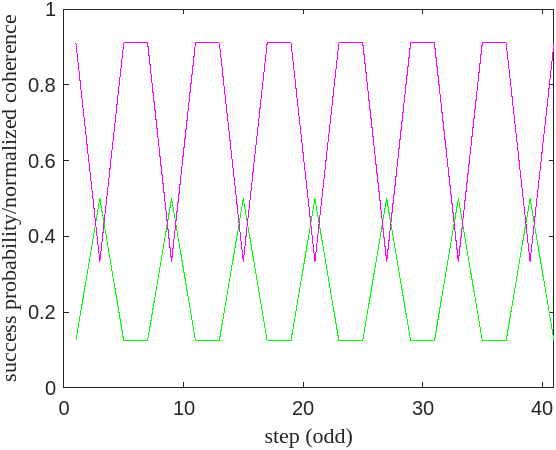}}
    \caption{The success probability and normalized coherence with the initial state $|s\rangle$ where $k=1$ and $N_{1}=N_{2}=4$. The blue and red {curves} are the success probability and normalized coherence for even $t$; The green and {purple curves} represent the success probability and normalized coherence when $t$ is an odd number respectively.}
    \label{c1}
\end{figure}

In order to show the evolution more clearly and intuitively, we demonstrate the relationship between success probability and normalized coherence by numerical simulations. When there exists one marked vertex and each partite set has four vertices, the dynamics of success probability and normalized coherence during the algorithm progress are shown in Figure \ref{c1}. The blue and red {curves} represent the success probability and normalized coherence when the time $t$ is even, respectively. The green and {purple curves} correspond to the case when $t$ is an odd number, which can be obtained by just shifting the even case and this is also consistent with the above theoretical derivation. 
It is not hard to find that the maximum value of the success probability corresponds to the minimum value of the normalized coherence, and vice versa. Also there is a complementary relationship, i.e. the sum of success probability and normalized coherence is approximately equal to 1.  
In fact, the Grover algorithm, which can be seen as quantum walk search on complete graph, has the similar behavior \cite{rastegin2018degradation, pan2022complementarity}. Thus, from this point of view, the quantum walk search on complete graph and complete bipartite graph are consistent.

{Moreover, we can obtain the complementary relationship between the success probability and normalized coherence in the asymptotic sense by the following analysis for all the two case.}
For the initial state $|s\rangle$, when $t$ is an even number, we have
\begin{align*}P_{t}+C_{l_{1}}(|\psi(t)\rangle)&=
   \cos^{2}{\theta}sin^{2}{A}+\sin{2\theta}(|\sin{A}|+|\cos{A}|)(|\sin{B}|+|\cos{B}|)+\cos^{2}{\theta}|\sin{2A}|+\sin^{2}{\theta}|\sin{2B}|\\
   &\approx\cos^{2}{\theta}sin^{2}{x}+\sin{2\theta}(|\sin{x}|+|\cos{x}|)^{2}+|\sin{2x}|\\
   &=\cos^{2}{\theta}sin^{2}{x}+\sin{2\theta}(1+|\sin{2x}|)+|\sin{2x}|\\ &=\cos^{2}{\theta}sin^{2}{x}+(1+\sin{2\theta})|\sin{2x}|+\sin{2\theta}\\
   &\approx\sin{2\theta},
\end{align*}
where $x=t\phi$. The above approximations are based on the fact that $\delta$ and $\phi$ approach 0 as $N_{i} (i=1,2)$ approaches infinity. The same property also holds when $t$ is odd.
Similarly, for the initial state $|\sigma\rangle$ and even number $t$, we have
\begin{align*}\tilde{P}_{t}+C_{l_{1}}(|\tilde{\psi}(t)\rangle)&=
   \frac{1}{2}\sin^{2}{A}+(|\sin{A}|+|\cos{A}|)(|\sin{B}|+|\cos{B}|)+\frac{|\sin{2A}|+|\sin{2B}|}{2}\\
   &\approx\frac{1}{2}\sin^{2}{x}+(|\sin{x}|+|\cos{x}|)^{2}+|\sin{2x}|\\
   &=\frac{1}{2}\sin^{2}{x}+2|\sin{2x}|+1\\ 
   &\approx1.
\end{align*}
And we can prove the property when $t$ is an odd number in the similar way.

\begin{figure}[htbp]
    \centering    
    \subfigure{\includegraphics[width=63mm]{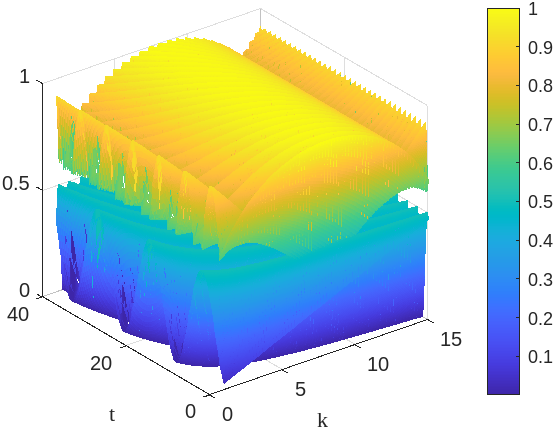}} 
    \subfigure{\includegraphics[width=53mm]{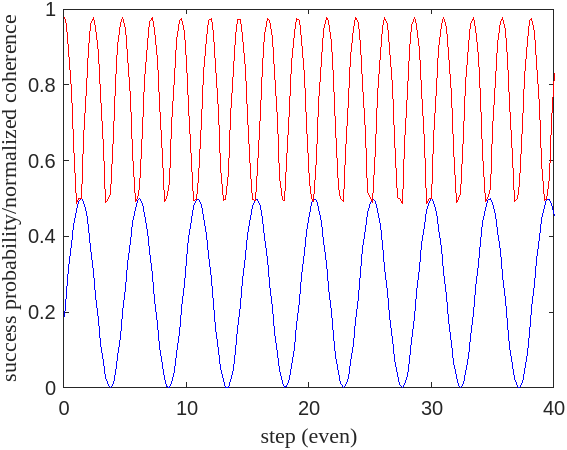}}
    \subfigure{\includegraphics[width=53mm]{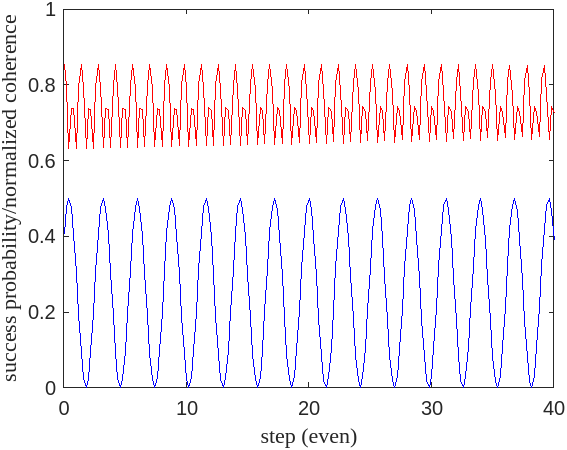}}
    \caption{{The first subfigure shows} the success probability and normalized coherence with different step $t$ and the number of marked vertices $k$, where the initial state is $|s\rangle$, $N_{1}=N_{2}=16$ and $t$ is an even number. {The upper and lower surfaces represent the normalized coherence and success probability respectively. The latter two sunfigures show two situations when the number of marked vertices $k$ is 6 and 13, where the blue and red curves are the success probability and normalized coherence, respectively.}}
    \label{c2}
\end{figure}

Furthermore, we also consider the evolution of the success probability and normalized coherence as functions of the number of steps and marked vertices. To be specific, we take the case that each partite set has sixteen vertices as an example, which is shown in Figure \ref{c2}. As we can see from the picture, regardless of the number of marked vertices, the correspondence between maximum and minimum value of success probability and normalized coherence holds, and the complementary relationship also holds.

\subsection{Dynamics of Entanglement in the Algorithm Progress}\label{ssec2}
\par 
\par Entanglement is an important resource in quantum information theory \cite{horodecki2009quantum,chitambar2019quantum}. Recently the roles that entanglement played in quantum algorithms attract much attention \cite{shi2017coherence,pan2019entangling,jaffali2019quantum,naseri2022entanglement}. In this subsection, we will study the effects of entanglement during the search algorithm on the complete bipartite graph based on the bipartite and multipartite entanglement measures. \par
First we recall a usual entanglement measure for bipartite systems, concurrence. For a bipartite pure state $\ket{\psi}_{AB}\in \mathcal{H}_A\otimes\mathcal{H}_B$, its concurrence is defined as \cite{hill1997entanglement,wootters1998entanglement}
\begin{align}
C(\ket{\psi}_{AB})=\sqrt{2(1-\tr{\rho_A^2)}},\label{c}
\end{align}
here $\rho_A$ is the reduced density matrix of $\ket{\psi}_{AB}.$ When $\rho_{AB}$ is a mixed state, its concurrence is defined as
\begin{align*}
C(\r_{AB})=\min_{\{p_i,\ket{\phi_i}_{AB}\}}\sum_i p_iC(\ket{\phi_i}_{AB}),
\end{align*}
here the minimum takes over all the decompositions of $\rho=\sum_ip_i\ket{\phi_i}_{AB}\bra{\phi_i}.$ When $\rho_{AB}$ is a two-qubit state, its concurrence can be computed as follows \cite{wootters1998entanglement},
\begin{align}
C(\rho_{AB})=\max(0,\lambda_1-\lambda_2-\lambda_3-\lambda_4),\label{cc}
\end{align}
where $\lambda_i$ are the square roots of the eigenvalues of  $\rho\tilde{\rho}$ in decreasing order, here $\tilde{\rho}=(\sigma_2\otimes\sigma_2)\rho^{*}(\sigma_2\otimes\sigma_2),$ $\sigma_2=\begin{pmatrix}
0&-i\\i&0
\end{pmatrix}$, $\rho^{*}$ is the conjugation of $\rho.$ Next we denote $sC(\ket{\psi}_{A_1A_2\cdots A_n})$ as the sum of concurrence for all the two-qubit reduced density matrices of $\ket{\psi}_{A_1A_2\cdots A_n},$
\begin{align}
    sC(\ket{\psi}_{A_1A_2\cdots A_n})=\sum_{i<j}C(\rho_{ij}),\label{ac}
\end{align}
here $\rho_{ij}$ is the reduced density matrix of $\ket{\psi}.$\par
Next we recall an entanglement measure for n-qubit pure states. In \cite{carvalho2004decoherence}, the concurrence for an $n$-qubit state $\ket{\psi}$ is defined as
\begin{align}
MC(\ket{\psi})=2\sqrt{\frac{2^n-2-\sum_{\alpha} Tr\rho_{\alpha}^2}{2^{n}}}, \label{con}
\end{align}
where $\alpha$ labels all the different reduced density matrices.

\indent In the subsection, we will present the roles entanglement plays in a simple case of this algorithm which is shown in Fig. \ref{1} based on the bipartite entanglement measure (\ref{ac}) and multipartite entanglement measure (\ref{con}). Assume the algorithm is executed in an $n$-qubit system and there exists only one marked vertex. As this search algorithm is on the complete bipartite graph, there should exist a qubit to mark the information on which partite the state is. A straightforward method is to make the first qubit to label the vertex in set $X$ or $Y$, that is, when the state is located in the set $X$ of the complete bipartite graph, the first qubit is labeled as $\ket{0}$, otherwise, the first qubit is labeled as $\ket{1}$. The remaining $n-1$ qubits are used to encode the information of vertices in set $X$ or $Y.$ And we can always assume the marked vertex is in set $X$ which is labeled as $\ket{00\cdots0}$.
As shown in Fig. \ref{1}, the states in the algorithm is in the space   $span\{\ket{ab},\ket{ba},\ket{bc},\ket{cb}\},$ and the four pure states can be written as 
\begin{align*}
\ket{ab}=&\frac{1}{\sqrt{2^{n-1}}}\ket{0}\ket{\underbrace{00\cdots0}_{n-1}}\ket{1}(\ket{\underbrace{00\cdots00}_{n-1}}+\ket{\phi}),\nonumber\\
{\ket{ba}}=&\frac{1}{\sqrt{2^{n-1}}}\ket{1}(\ket{\underbrace{00\cdots00}_{n-1}}+\ket{\phi})\ket{0}\ket{\underbrace{00\cdots0}_{n-1}},\\
\ket{bc}=&\frac{1}{\sqrt{2^{n-1}(2^{n-1}-1)}}\ket{1}(\ket{\underbrace{00\cdots00}_{n-1}}+\ket{\phi})\ket{0}\ket{\phi},\\
\ket{cb}=&\frac{1}{\sqrt{2^{n-1}(2^{n-1}-1)}}\ket{0}\ket{\phi}\ket{1}(\ket{\underbrace{00\cdots00}_{n-1}}+\ket{\phi}),
\end{align*}
where $\ket{\phi}=\ket{\underbrace{00\cdots01}_{n-1}}+\ket{\underbrace{00\cdots10}_{n-1}}+\cdots+\ket{\underbrace{11\cdots11}_{n-1}}.$ The states in the algorithm can always be written as \begin{align*}
\ket{\psi(t)}=m_1\ket{ab}+m_2\ket{ba}+m_3\ket{bc}+m_4\ket{cb},
\end{align*}
here $m_i$ $(i=1,2,3,4)$ are functions on $t$.
\par 
 Based on (\ref{ac}), we have the sum of entanglement for all the bipartite states of $\ket{\psi(t)}$ is

\begin{align} 
sC(\ket{\psi(t)})=&\max(\frac{m_1m_2}{2^{n-1}}+\frac{\sqrt{2^{n-1}-1}}{2^{n-1}}(m_1m_3+m_2m_4)+\frac{2^{n-1}-1}{2^{n-1}}m_3m_4,0).\label{sc}
\end{align}\par
\begin{figure}
\centering
   \subfigure{\includegraphics[width=89mm]{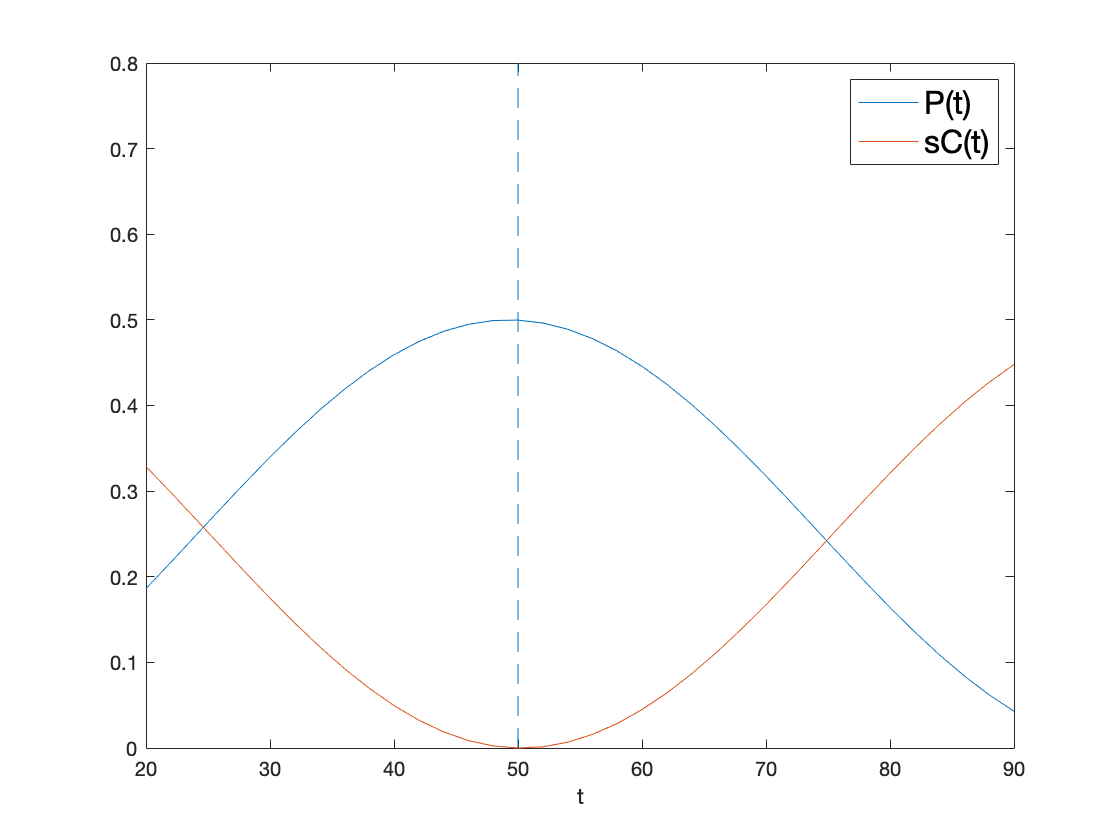}}
    \subfigure{\includegraphics[width=89mm]{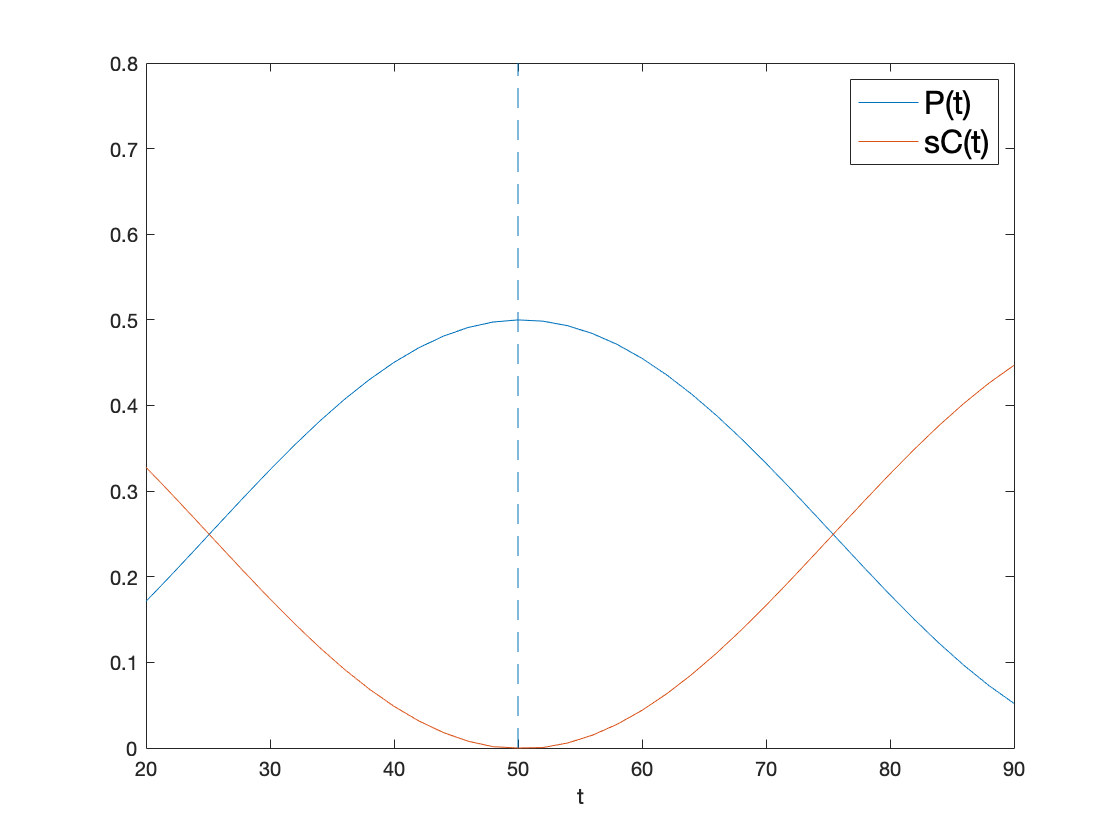}}
 \caption{The success probability and the sum of the concurrence for all the bipartite reduced density matrices with one marked vertex when $n=10$. }\label{fi1}
     \end{figure}
\begin{figure}
\centering \subfigure{\includegraphics[width=89mm]{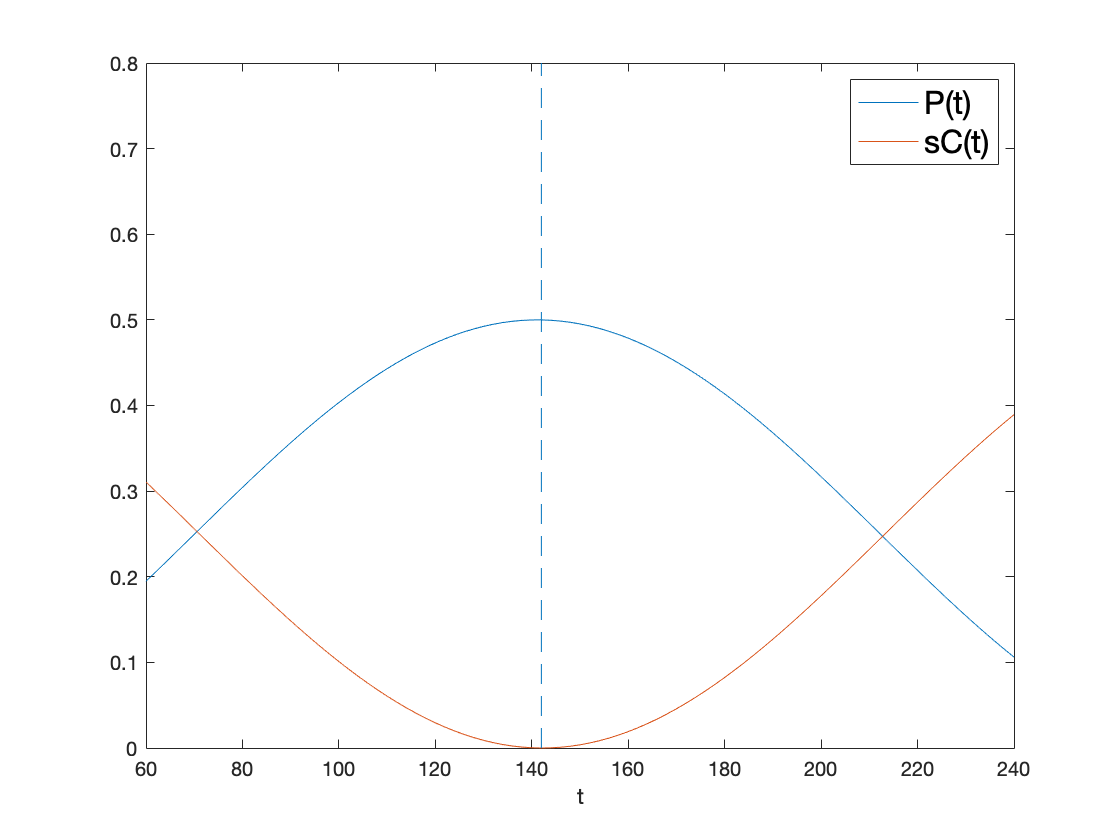}}
    \subfigure{\includegraphics[width=89mm]{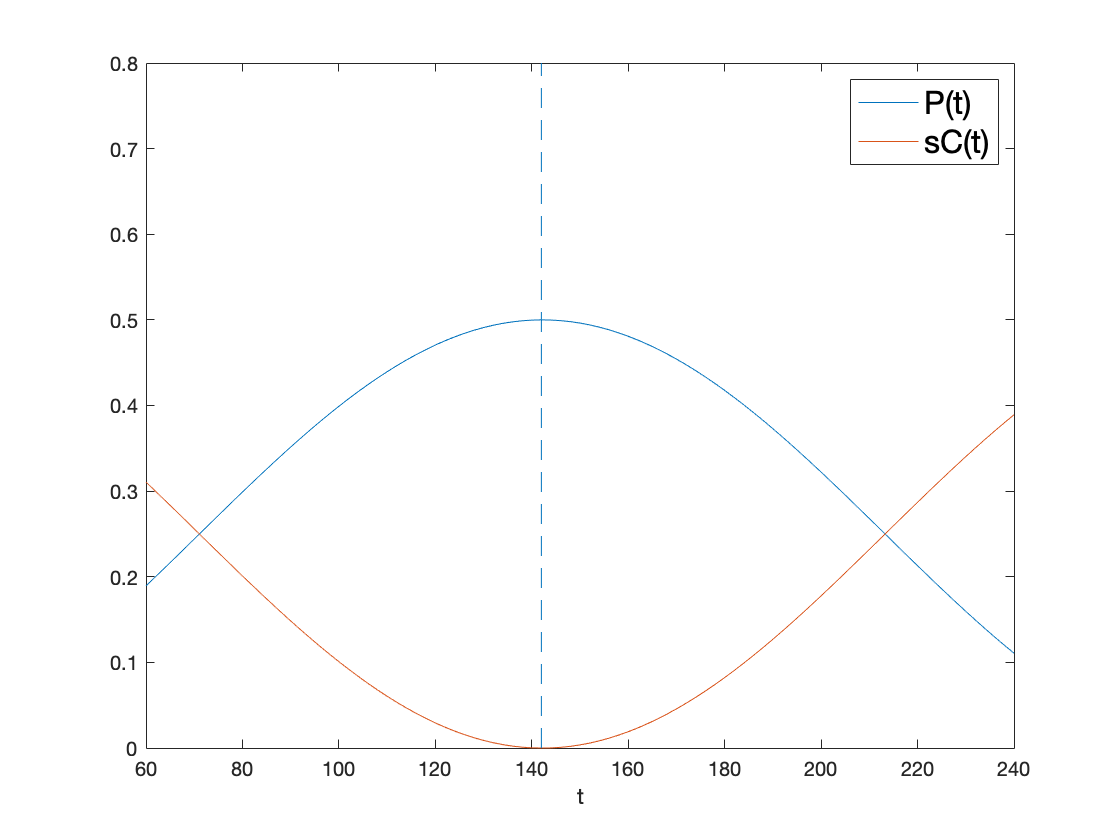}}
    \caption{The success probability and the sum of the concurrence for all the bipartite reduced density matrices with one marked vertex when $n=13$. }\label{fi2}
     \end{figure}

Next we denote $sC(\ket{\psi(t)})$ as $sC(t)$ in order to simplify the expression and be consistent with the success probability $P(t)$. And we present the relationship between $P(t)$ and  $sC(t)$ with the increasing steps in Fig. \ref{fi1} and Fig. \ref{fi2}. There the two pictures in Fig. \ref{fi1} and Fig. \ref{fi2} are on the cases when $N_1=N_2=2^{10}$ and $N_1=N_2=2^{13},$ respectively. At the same time, the left are on the cases when $t$ is even, while the right are on the cases when $t$ is odd. And in the four pictures, the blue {curve} $x=P(t)$ represents the success probability, while the orange {curve} $z=sC(t)$ represents the sum of entanglement for all the bipartite reduced density matrices. Due to the four pictures, we have when $P(t)$ reaches the maximum, $sC(t)$ reaches the minimum. Therefore, we may conclude that the entanglement plays important roles in the search algorithm on the complete bipartite graph.
\par 
 Then we will compare the role entanglement plays here with that in GSA. In \cite{shi2017coherence},  the authors presented that the relations between $sC(t)$ and $P(t)$ in GSA seem irrelevant. Nevertheless, the relation between $sC(t)$ and $P(t)$ in the search algorithm here is complementary from Fig. \ref{fi1} and Fig. \ref{fi2}. Specifically, when the success probability reaches the maximum, $sC(\cdot)$ arrives at the minimum. Hence we may conclude that this complementary relation is closely related to the search algorithm on different topological structures.  
\par
{Furthermore, we can obtain the complementary relation between $P(t)$ and $sC(t)$ in the asymptotic scenario analytically of the cases considered here. When $n$ tends to infinity and $t$ is even, 
\begin{align*}
&\frac{m_1m_2}{2^{n-1}}+\frac{\sqrt{2^{n-1}-1}}{2^{n-1}}(m_1m_3+m_2m_4)+\frac{2^{n-1}-1}{2^{n-1}}m_3m_4\nonumber\\
    =&\frac{\sqrt{2^{n-1}-1}}{2^{n-1}}\sin\theta\cos\theta\sin2\delta+\frac{\sin\theta\cos\theta\sin(\delta+\phi t)\sin(\delta-\phi t)+(2^{n-1}-1)(\sin\theta\cos\theta\cos(\delta+\phi t)\cos(\delta-\phi t)}{2^{n-1}}\nonumber\\
    \approx&\cos^2\theta\cos^2(\phi t+\delta)+2\cos^2\theta\cos(\phi t+\delta)\sin(\phi t)\sin\delta\\
    \approx&\cos^2\theta\cos^2(\phi t+\delta),
\end{align*}
here $\theta,\delta$ and $\phi$ are defined below $(\ref{sp})$. And the last two are due to that when $n\rightarrow \infty,$ $\delta,\phi\rightarrow 0.$ As $P(t)=\cos^2\theta\sin^2(\delta+\phi t),$ $P(t)+sC(t)\approx \cos^2\theta=\frac{1}{2}.$ Similarly, we can show the property when $t$ is odd.}\par
\begin{figure}[htbp]
    \centering  \includegraphics[width=0.55\textwidth,height=0.35\textwidth]{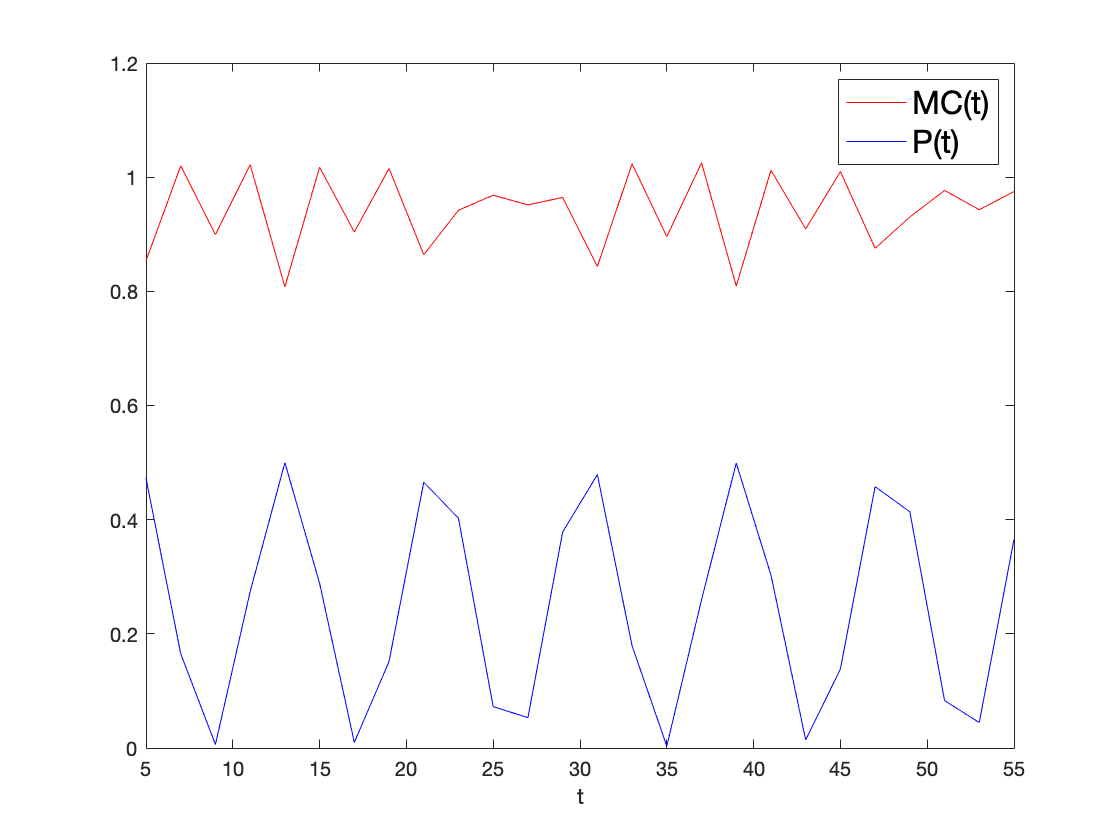}
    \caption{Success probability and multipartite entanglement for search on the complete bipartite graph when $N_1=N_2=2^3$ and t is odd. The blue {curve} and the red {curve} are on the success probability and the concurrence of the state under evolution, respectively.}
    \label{f4}
\end{figure}

 
At last, let us consider the multipartite entanglement of the state in the algorithm process.  In Fig. \ref{f4}, we plot the behaviors of multipartite entanglement and its success probability of the states in the algorithm when the number of the vertices in set $X$ and $Y$ are $2^3$. By contrast, the quantity on bipartite entanglement measures owns a clearer relation with the success probability of the algorithm, hence it may be seen as a more meaningful indicator on the algorithm. And this fact can be explained as follows, as the labeled state is $\ket{00\cdots0},$ when the success probability of this algorithm gets the biggest, the distance between the states of the algorithm and the labeled state is the closest, hence the two-qubit reduced density matrices of the states in the algorithm are the nearest to the set of separable states. However, due to the existence of genuine entanglement, the above fact may be invalid for multipartite entanglement.
\par 

\section{Dynamics of Quantum Coherence Under Noises}\label{ssec3}
Real noise is inevitable in the practical implementation of a specific quantum evolution process. There are various distinct noise channels, such as amplitude damping, phase damping and depolarization. In particular, the depolarizing noise channel, as an important type of noise, is popular to discuss quantum error-correction.
The general quantum depolarizing operation on a $d$-dimensional system is $$E(D)=\frac{1-\alpha}{d}Tr(D)I_{d}+\alpha D.$$ 
For the normalized initial state $|\psi(0)\rangle$, the corresponding density matrix, success probability and coherence at $i$-step quantum walk under general quantum depolarizing noises can be described as {$\delta_{i}$}, $Q_{i}$ and {$C_{l_{1}}(\delta_{i})$}, respectively, where ${\delta_{0}}=|\psi(0)\rangle\langle\psi(0)|$. Thus, it is straightforward to deduce:
$${\delta_{1}}=UE({\delta_{0}})U^{\dagger}=U(\frac{1-\alpha}{4}I+\alpha{\delta_{0}})U^{\dagger}=\frac{1-\alpha}{4}I+\alpha U{\delta_{0}}U^{\dagger},$$
$${\delta_{2}}=UE({\delta_{1}})U^{\dagger}=U(\frac{1-\alpha}{4}I+\alpha{\delta_{1}})U^{\dagger}=\frac{1-\alpha}{4}(1+\alpha)I+\alpha^{2}U^{2}{\delta_{0}}(U^{\dagger})^{2},$$
$$\cdots$$
$${\delta_{n}}=UE({\delta_{n-1}
})U^{\dagger}=U(\frac{1-\alpha}{4}I+\alpha{\delta_{n-1}})U^{\dagger}=\frac{1-\alpha^{n}}{4}I+\alpha^{n}U^{n}{\delta_{0}}(U^{\dagger})^{n}.$$

\begin{figure}[htbp]
    \centering   \includegraphics[width=0.45\textwidth,height=0.28\textwidth]{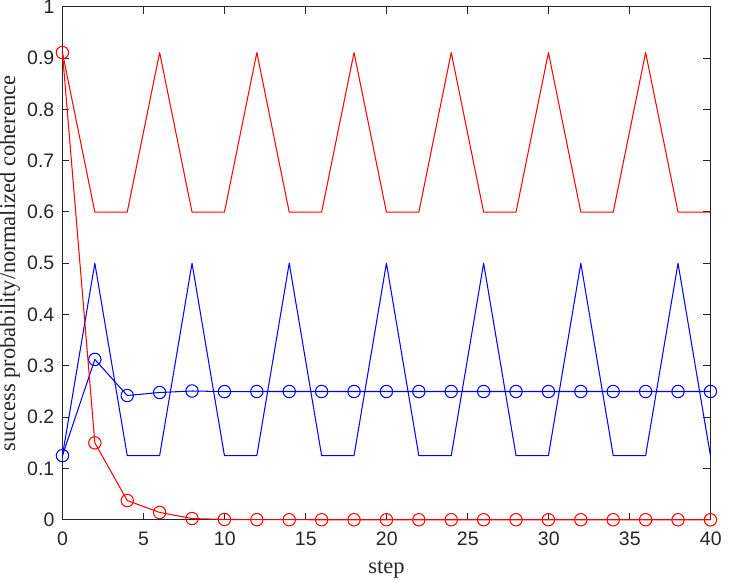}
    \caption{The success probability and normalized coherence with the initial state $|s\rangle$ where $k=1$, $N_{1}=N_{2}=4$ and $t$ is an even number. The blue and red represent the success probability and the normalized coherence. The solid and dashed {curves} correspond to the noiseless and noisy cases. The parameter in general quantum depolarizing operation here is $0.5$.}
    \label{c3}
\end{figure}

To analyze the effect of noise, we denote the density matrix, success probability and coherence at $i$-step quantum walk with no noises as $\rho_{i}$, $P_{i}$ and $C_{l_{1}}(\rho_{i})$, respectively. 
Note that $\rho_{0}=\sigma_{0}$ and the state after $n$-step quantum walk is $\rho_{n}=U^{n}\rho_{0}(U^{\dagger})^{n}$.
According to the definition of success probability and $l_{1}$-norm coherence, we can obtain
$$Q_{n}=\frac{1-\alpha^{n}}{4}+\alpha^{n}P_{n},\ \ \  C_{l_{1}}({\delta_{n}})=\alpha^{n}C_{l_{1}}(\rho_{n}).$$
Next, we demonstrate the effect of the general depolarizing operation directly by taking numerical simulation in Figure \ref{c3}. Here we only discuss the case when $\alpha=0.5$. And we can see the coherence decreases dramatically and turn to zero after the first few steps. Compared with the case without noise, the success probability is reduced and gradually become a fixed value. Also, success probability and coherence are no longer periodic.

\section{Summary}\label{sum}
In this paper, we mainly studied the evolution of quantum coherence and quantum entanglement in the quantum walk on the complete bipartite graph. On the one hand, there is a correspondence between the maximum (minimum) of quantum resources and the minimum (maximum) of success probability. Furthermore, we proved that there exists a strong complementary property between success probability and quantum coherence (entanglement) in the asymptotic sense. On the other hand, we discovered that the generalized depolarizing noises can influence the quantum coherence and performance of the search algorithm. We believe our work will deepen our understanding of quantum resources and inspire us to design more efficient quantum algorithms.

\section{Acknowledgments}
This work was supported by China Postdoctoral Science Foundation (Grant No.2022M723209), the National Natural Science Foundation of China (Grant No.62301531 and No.12301580), the Fundamental Research Funds for the Central Universities (Grant No.ZY2306), and Funds of College of Information Science and Technology, Beijing University of Chemical Technology (Grant No.0104/11170044115).

\bibliographystyle{unsrt}
\bibliography{propertyref}

\end{document}